# Quantum-Coherent Nanoscience

Andreas J. Heinrich[1,2,*], William D. Oliver[3,4], Lieven Vandersypen[5], Arzhang Ardavan[6], Roberta Sessoli[7], Daniel Loss[8], Ania Bleszynski Jayich[9], Joaquin Fernandez-Rossier[10,11], Arne Laucht[12], and Andrea Morello[12,*]

1. Center for Quantum Nanoscience (QNS), Institute for Basic Science, Seoul 03760, Korea
2. Physics Department, Ewha Womans University, Seoul 03760, Korea
3. MIT Research Laboratory of Electronics, 77 Massachusetts Avenue, Cambridge, MA 02139, USA
4. MIT Lincoln Laboratory, 244 Wood Street, Lexington, MA 02421, USA
5. QuTech and Kavli Institute of Nanoscience, TU Delft, Lorentzweg 1, 2628CJ Delft, The Netherlands
6. CAESR, The Clarendon Laboratory, Department of Physics, University of Oxford, Parks Road, Oxford OX1 3PU, UK
7. Department of Chemistry "U. Schiff" & INSTM, University of Florence, 50109 Sesto Fiorentino, Italy
8. Department of Physics, University of Basel, Klingelbergstrasse 82, 4056 Basel, Switzerland
9. Department of Physics, University of California Santa Barbara, CA 93106, USA
10. QuantaLab, International Iberian Nanotechnology Laboratory (INL), Avenida Mestre José Veiga, 4715-310 Braga, Portugal
11. Departamento de Física Aplicada, Universidad de Alicante, San Vicente del Raspeig 03690, Spain
12. School of Electrical Engineering and Telecommunications, UNSW Sydney, NSW2052, Australia

* Email: AJH: heinrich.andreas@qns.science, AM: a.morello@unsw.edu.au

## Abstract

For the past three decades, nanoscience has widely affected many areas in physics, chemistry, and engineering, and has led to numerous fundamental discoveries as well as applications and products. Concurrently, quantum science and technology has developed into a cross-disciplinary research endeavour connecting these same areas and holds a burgeoning commercial promise. Although quantum physics dictates the behaviour of nanoscale objects, quantum coherence, which is central to quantum information, communication and sensing has not played an explicit role in much of nanoscience. This Review describes fundamental principles and practical applications of quantum coherence in nanoscale systems, a research area we call quantum-coherent nanoscience. We structure this manuscript according to specific degrees of freedom that can be quantum-coherently controlled in a given nanoscale system such as charge, spin, mechanical motion, and photons. We review the current state of the art and focus on outstanding challenges and opportunities unlocked by the merging of nanoscience and coherent quantum operations.

Quantum mechanics dictates the basic laws that govern the behaviour of matter and fields at the scale of particles, atoms, and molecules. Therefore, all functionalities at the nanoscale are, at some level, underpinned by quantum effects. Nevertheless, many current functional nanoscale devices are designed and operated without resorting to an explicit quantum treatment. In this manuscript, we describe a more recent development in nanoscience, where quantum coherent phenomena at the nanoscale are *explicitly* harnessed to enable novel functionalities. We call this endeavour quantum-coherent nanoscience. It represents the core capability that will empower emerging quantum technologies such as quantum computing, quantum simulation, quantum communication, and quantum sensing based on solid-state or molecular nanosystems. In quantum-coherent nanoscience, the behaviours and functionalities that one seeks are the profound, fundamental properties of quantum states, such as superposition, entanglement, and quantum coherence. We will describe how such quantum properties can be designed to persist in engineered materials and devices at the nanoscale. Quantum-coherent nanoscience, lying at the intersection between nanoscience and quantum science, is a discipline within the fields of condensed matter physics, materials science, and molecular chemistry.

There are three broad reasons for exploring the nanoscale. First, a lesson from the development of classical information technologies is the value of miniaturisation. A quantum technology built on nanoscience has the potential for a density of components approaching the exceptional scale achieved by semiconductor devices, which has enabled the modern era of information technology.

Second, a range of useful phenomena occur when available quantum excitations are confined or manipulated at the nanoscale, see Box.1 . For example, electrons confined to regions in a semiconductor with dimensions of order tens of nanometers find their kinetic energy quantised on the scale of $k_B$*10 K, offering the possibility of engineering quantum dots with tailored properties[1]. In high-dielectric materials, structures with features on a similar length scale (photonic crystals) can be used to tailor, guide, and localise electromagnetic excitations[2]. Additionally, nanoscale objects constructed with specific materials exhibit acoustic modes with frequencies in the GHz range[3,4], allowing the control of individual quanta of mechanical motion.

Third, if the quantum properties of these excitations are to be exploited, their interactions with the environment and with each other must be controlled, which in many systems leads to an associated length scales in the nanometer range. For example, the tunnelling of electrons between neighbouring quantum dots separated by a nanoscale barrier can be sufficient to create and manipulate coherent superpositions of charge states[5]. In combination with the Pauli exclusion principle, the analogous tunnelling of electrons between atoms positioned with few lattice spacing separations on a surface[6] or inside a material[7], or through the chemical bonds in molecules[8], gives rise to effective exchange or dipolar magnetic interactions which offer the means to generate entanglement between spins[9,10,11].

The goal of this Review is to illustrate how the ability to design and control structures at the nanoscale enables the implementation of quantum coherent functionalities. We do so by describing a non-exhaustive set of practical examples, organized on the basis of the physical degrees of freedom (DOF) involved in the quantum dynamics: electrical charge in quantum dots; spins of electrons and nuclei associated with quantum dots, defects in insulators and semiconductors, atoms on surfaces, and magnetic molecules; photons emitted and detected by nanostructures; nanomechanical oscillators; and hybrid quantum systems involving coherent coupling between different DOFs. We note that quantum coherence is a natural and crucial aspect of the behaviour of trapped ions and cold atomic gases. However, since these systems do not require engineered nanostructures in order to harness their quantum properties, we shall not cover them in this nanoscience review.

## Manipulating quantum states of electrical charge at the nanoscale

The degree of freedom of the electrical charge is undoubtedly the most crucial for today's information technologies, ranging from transistors for data processing to flash memories for long-term data storage. Therefore, we start our investigation into quantum-coherent nanoscience with the charge degree of freedom.

Silicon transistor technology has fuelled the information processing revolution of the last fifty years, driven in large part by a reduction in device dimension that reached the realm of nanoscience (< 100 nm) in commercial products in the early 2000's. The industry is now at the 7 nm node, producing devices with critical dimensions in the 10-40 nm range. At these length scales, confined electrons begin to exhibit quantum effects. While the transistors in today's processors are carefully designed to work around such quantum effects and operate as classical logic elements, the commercialization of electronic devices in the nanoscale creates future opportunities for fundamentally quantum-mechanical electronic devices.

Quantum dots in a semiconductor comprise electrons confined to nanoscale islands[1,12,13]. The strong confinement and Coulomb interaction limit the number of electrons on the island – down to even a single electron. At cryogenic temperatures, the confined electron(s) exhibit discrete orbitals along with the electron spin degree of freedom. Due to their resemblance to natural atoms, quantum dot systems are often described as artificial atoms[1,13]. Charging effects are similarly observed in superconducting islands, where pairs of electrons called Cooper pairs can be added one at a time.

Both individual electrons and Cooper pairs can exist in spatial superpositions on two separate islands[14,15] and coherently oscillate between them, forming the basis for a charge qubit[5,16], see Fig.1. Charge qubits have featured energy decay times exceeding 100 µs [17], but due to ubiquitous charge noise, the dephasing is generally limited to below 1 µs with pronounced temporal and device-to-device variations. However, this strong interaction with the environment also enables charge qubits to become strongly coupled to other quantum degrees of freedom, resulting in excellent sensors as well as hybrid quantum systems[18,19]. For instance, the addition of a single electron charge to a quantum dot or donor is easily detected by a nearby electrometer and, in combination with spin-dependent tunnelling, also facilitates single-shot spin readout[20,21]. Redesigning superconducting charge qubits to suppress the impact of charge noise[22] – such that the phase rather than the charge degree of freedom dominates the behaviour – has reproducibly improved qubit coherence times of superconducting devices beyond 100 µs with gate times in the 10-50 ns range[23]. Recently, highly precise control of a circuit of 53 superconducting qubits was demonstrated; at this scale it is no longer feasible for a classical supercomputer to simulate the qubits' state[24].

## Magnetic nanosystems: The spin degree of freedom

The achievement of quantum-coherent manipulation and readout of individual spins in solids and molecules reinforces the promise of condensed-matter spins for applications in quantum information and quantum sensing. In terms of international research efforts, the coherent manipulation of spins comprises the largest part of quantum-coherent nanoscience. Systems currently under investigation include magnetic defects in insulators[25,26] and semiconductors[21,27,28], spins in lithographically-defined quantum dots[29,30,31], magnetic molecules[32,33,34], and magnetic atoms on surfaces[6,35,36]. The intrinsic compactness of spin-based devices holds promise for dense integrated technologies and plays an important role in spatially-resolved quantum sensing. Conversely, the small inherent length scale results in challenges for controlling individual qubits and their interactions in quantum information technologies.

Magnetic resonance provides a well-established framework for manipulating the quantum state of both electronic and nuclear spins. Nevertheless, a surprising variety of implementations based on oscillating magnetic or electric fields[12,13] as well as photonic sources[37] have been employed, see Box 2. Confining individual nuclear and electron spins such that they can be individually addressed as qubits (a term which we do not limit to quantum computation) requires devices with nanoscale dimensions.

In addition to the formation of qubits and control of quantum-coherent states, the nanoscale plays a key role in the readout of spin states, which often requires integration in nanoscale electronic devices. This is particularly true for spin to charge conversion in quantum dots, dopants in semiconductors, and scanning tunnelling microscopy. In special cases, optical excitations and detection of single-spin states are possible, for example in the case of colour centres, e.g. nitrogen-vacancy in diamond[25,26,38], or optically-active spin centres in silicon carbide[28] and in molecules designed for this purpose[34].

## Artificially Assembled Nanostructures

Quantum dots offer a quintessential example of a spin in a nanoscale system since an individual confined electron in such an artificial atom has a spin of $S = ½$ [29]. Alternatively, multiple electrons in different dots can be combined to have two-level quantum spaces amenable to electrical control[39,40]. The spin is much less sensitive to its microscopic environment than the charge qubits[13,30,31], and coherence times up to tens of milliseconds[41] have been reported when using host materials such as $^{28}$Si, which contains no nuclear spins. Using this platform, two-qubit logic gates[42] and simple two-qubit algorithms have already been implemented[43], linear arrays of nine quantum dots can be controllably formed[44] and universal quantum logic with fidelity above 99% [45], and entanglement between three electron spins [46] have been demonstrated. Four qubits in a 2x2 array have been controlled in a germanium-based platform [47]. Figure 2a,b shows an example of creating and controlling entangled states in quantum-dot spins, a key feature that unambiguously distinguishes quantum from classical behaviour[43].

Lithographically defined nanostructures also play a key role when controlling the quantum properties of individual dopants in a semiconductor. Typically, their spin degree of freedom is only weakly coupled to other excitations, making such spin qubits highly coherent. In isotopically-enriched $^{28}$Si, dilute bulk $^{31}$P donor electron spins can remain coherent for 10 seconds, and the nuclear spin for 3 hours[48]. In functional nanoelectronic devices with the infrastructure for qubit manipulation and readout, these values remain close to 1 second for the electron and 30 seconds for the nuclear spin[49]. Figure 2c,d shows such an example and illustrates that a two-qubit system consisting of one electron and one nuclear spin can be used to unequivocally demonstrate entanglement between these two qubits by violating the Bell's inequality[10]. Such entanglement has been extended to a system of two nuclei and one electron, wherein nuclear two-qubit gate fidelities exceed 99% [50].

The challenges in the exploitation of donor spins in semiconductors for quantum information lie almost entirely in the fabrication. A scalable quantum processor requires large arrays of controllably coupled qubits, demanding very tight fabrication protocols, or creative ways to introduce multi-qubit interactions. These scalability challenges are being tackled using two alternative approaches. The first is to push the fabrication tolerances to the sub-nanometre level, using scanning probe lithography. This method yields extremely precise placement of the donors, which typically aggregate in small clusters of 2-3 atoms[7]. The second method continues to develop the industry-standard ion implantation method, which leaves an inherent placement uncertainty of order 10 nm, but seeks to encode and manipulate the quantum information in a way that makes the system less sensitive to exact donor position[51].

Quantum-coherent measurements of spin states of individual defects in insulators were one of the initial drivers of quantum-coherent nanoscience, with the Nitrogen-Vacancy (NV) centre in diamond as the most prominent early contestant[52]. Initialization and read-out of the low-energy $S = 1$ spin states is achieved via resonant or non-resonant optical excitation[52]. In contrast to most other spin systems discussed here, this is even possible at room-temperature with good quantum coherence times. Due to the random placement of the defects, studying isolated qubits requires high dilution. However, these qubits are often coupled to nearby nuclear spins offering quantum-coherent control of multi-partite systems[53]. NV centres and related defects have become powerful quantum sensors with nanometre-scale spatial and very high electric and magnetic field resolution, see Fig.3 c,d. Challenges in using NV centres as quantum sensors stem from the decay of the quantum coherence when they are brought close to a surface[54], a requirement for achieving spatial resolution below 10nm. Progress is made, for example by using ion-beam lithography to place individual point defects in controlled positions with about 10 nanometre precision[55].

Atoms on clean surfaces offer yet another path towards controlling individual spins in solids. Recently, GHz-frequency electric fields between the tip and the sample in a Scanning Tunnelling Microscope (STM) were used to drive CW [35,36,56] and pulsed ESR[6] of single atoms, see Fig. 3a. While the access to spins on the atomic scale as offered in scanning probe microscopy has great benefits, this approach also has major disadvantages, such as the presence of conduction electrons tunnel-coupled to the spin. Every electron, even those that are inherent in the operation of the STM, gives rise to a loss of quantum coherence[57]. Thus, the quantum coherence time is presently limited to some 100ns (Fig. 3b) with gate operation times of about 10ns [6]. The detrimental effects of conduction electrons could be overcome by working with thick insulating films and very small tunnel currents.

Self-assembled Nanostructures

Self-assembly is the hallmark of chemistry, which yields parallel production of tuneable functional building blocks and integrated assemblies, which can be incorporated into nanoscale devices[32]. Molecular quantum nanoscience thus offers the enticing prospect of engineering structures to achieve intended functions at the smallest length scale possible. Early evidence of incoherent quantum tunnelling in molecular magnets stimulated theoretical and experimental work to push into the quantum coherent regime, exploiting magnetic molecules as components in quantum information technologies[8,58]. This motivated the design of magnetic molecules with coherence times of tens or hundreds[59] of microseconds at low temperatures and coherent spin dynamics persisting at room temperature[60]. Light[34] and electric fields[32,61] offer rapid manipulation of functional units and interconnects and may provide controls for implementing conditional classical or quantum information operations[62].

A major breakthrough came from measuring electrical transport on a single molecule incorporating a magnetic terbium atom, see Fig. 3e,f. Via a carefully crafted hierarchy of interactions between the itinerant electrons and the electronic and nuclear magnetic moments, it was possible to interrogate and manipulate the quantum state of the terbium nuclear spin[32] and thereby implement projective readout and simple quantum algorithms. The detrimental effect of conducting electrons is here reduced because they flow through the ligand and therefore only weakly perturb the metal ion's spin.

Spins in condensed matter systems are not only the building blocks of potential devices and technologies, but also underpin intriguing emergent quantum phenomena such as high-temperature superconductivity and spin liquids[63]. An outstanding challenge for quantum-coherent nanoscience is the fabrication of quantum simulators of 2D arrays of spins to simulate Hubbard models with a number of sites beyond the tractable limit of classical simulators, a limit that one could describe as Analog Quantum Advantage, in analogy with the quantum advantage term used in the context of digital quantum computing[24]. This is also an objective in ion and atom trap systems[64]. Steps in this direction have been taken, for instance, adatom spin chains realize 1D spin models[65,66], though these can still be efficiently simulated classically using density matrix renormalization methods. Quantum simulators of Hubbard dimers and 4-site plaquettes have been fabricated using dopants[67], spins on surfaces[68], and quantum dots[69], but the number of sites needs to increase to surpass simulations on classical computers.

# Nanoscale quantum-photonics

Photons have been a carrier of information for a long time – from the signal fires of ancient civilisations to the glass fibres of today's broadband internet – for very good reasons. Photons travel fast and have a high frequency, which allows high-rate information transfer. Furthermore, they largely do not interact, which makes them insensitive to electromagnetic interference and enables frequency multiplexing of data. In the age of quantum information, photons will have an even more important role to play: Quantum communication relies on the faithful emission and detection of single-photons[70]; quantum networks require stationary qubits to be entangled with single-photons[71, 72, 73]; and all-optical computers can operate much faster than their electronic counterparts when using optical transistors[74, 75]. The challenges are how to produce single-photons on-demand with high purity[76], how to detect them with high efficiency[77], and how to make them interact with either a stationary qubit[78, 79] or another photon[75]. To solve these challenges, we need to invoke nanoscale dimensions and quantum effects[80].

Optical photons have historically been produced by high temperatures, resulting in high numbers of photons with classical statistics that are unusable for quantum applications. The discrete energy levels in atoms, ions and molecules restrict photons to be emitted one-by-one, but trapping and holding a single atom or molecule can be a challenging task. Therefore, solid-state emitters – that are inherently stationary – have been studied as an alternative[81]. Atomistic defects in wide-bandgap semiconductors and 2D materials emit single-photons at high rates with low multi-photon probabilities[82, 83]. The currently best performing single-photon emitters are self-assembled InGaAs quantum dots (QDs) embedded in a GaAs host matrix[76]. The small dimensions of 10-30 nm in lateral and 2-5 nm in vertical direction quantize the energy levels and provide the atom-like emission spectrum required for single-photon emission. Integrating these QDs into photonic cavities to further enhance the spontaneous emission rate and improve the collection efficiency[76, 84], in combination with resonant optical excitation to impede multi-photon emission, has led to photon extraction efficiencies of 60-80%, single-photon purities of 99% and indistinguishability of 97-99% [76].

The detection of single-photons relies on converting their energy into a measurable electrical signal, usually via a cascade or avalanche effect as in photo-multiplier-tubes and avalanche photodiodes. The more recently developed superconducting-nanowire single-photon detectors outperform earlier detectors in terms of spectral bandwidth (ultraviolet to mid-infrared), temporal resolution (<10ps), dark counts (<10/s) and quantum efficiency (>90%)[77]. This is achieved using a superconducting nanowire of approximately 5x100nm cross-section that is biased close to the critical current. The absorption of a photon turns the wire into a normal conductor, raising its resistance and producing a measurable voltage drop. Different geometries and materials, optical cavities to optimize the absorption cross-section, and cryogenic amplifiers have been used to enhance the performance of these detectors and tailor them to specific applications. Further developments include on-chip integration on devices[85], photon number resolving detectors and 2D arrays[77].

Single-photon emission and detection rely on single emitters, quantum confinement, superconductivity and nanoscale dimensions, but they are predominantly based on incoherent effects. In order to create spin-photon entanglement or photonic switching, we do however require quantum-coherent interactions between photons, excitons and spins[80]. The challenge here is to enhance the photonic coupling beyond the losses and dissipation in the system. This is achieved by confining the photon to a photonic cavity with high cooperativity[79,86], or by tailoring the photonic dispersion relation to slow them down and prolong their interaction time[87]. Recent achievements include the photon-mediated interaction between two SiV centres in a diamond nanocavity that is switchable via the electronic spin state[78] and the heralded storage of a photonic qubit via spin-photon entanglement in a similar system[79]. For semiconductor spin qubits, the coherent coupling to microwave photons in superconducting cavities is similarly advanced[88]. Finally, there is photon switching[75,78], where a single-photon controls the transmission of a second incident photon through the coherent interaction of a quantum two-level system with a photonic cavity[75]. This allows the construction of optical transistors for building an all-optical computer[74].

The field of nanoscale quantum optics has seen huge progress over the last two decades[80]. Engineering of materials to supply bright, stable and coherent quantum emitters was as crucial as the availability of nano-patterning tools[72]. However, some of the current challenges on the way to realizing a quantum network are still related to samples and materials[71]. For example, state-of-the-art single-photon sources operate at near-infrared wavelengths just below 1μm, outside the telecom bands. QDs emitting at 1.3μm and 1.55μm exist but do not yet have the same performance, and wavelength conversion involves a loss in efficiency[76,81]. At the same time, material defects, surface roughness and fabrication defects can lead to enhanced dissipation and scattering of photons inside cavities and waveguides. This reduces the cooperativity of the quantum system – photon interaction, which is the key figure of merit for achieving coherent spin-photon coupling and photon-mediated spin-spin coupling as required for repeater nodes of quantum networks[71,78,79]. With increasing knowledge about material systems and improvements in fabrication processes these challenges can surely be tackled. Recent highly-integrated devices show us a glimpse of what will be possible in the future[72,85].

## Quantum limits of mechanical motion

Mechanical degrees of freedom, despite the participation of large numbers of atoms, often more than $10^6$, can now routinely be observed and operated in the quantum regime. The quantum-coherent control of phononic excitations is of interest for many reasons, including fundamental studies of quantum mechanics and decoherence at increasingly large scales, quantum memories, quantum sensors, and quantum transducers. Their broad applicability in the field of quantum science is enabled by a combination of their extraordinarily high quality factors and versatile coupling to a variety of fields.

Mechanical oscillators are ubiquitous in nature and in today's technologies, fuelling *e.g.* signal filtering in cell phones and gravitational wave sensing in the Laser Interferometer Gravitational-Wave Observatory (LIGO). With their exquisite sensitivity[89,90,91] for force of aN/Hz$^{1/2}$ and mass of yg/Hz$^{1/2}$, mechanical sensors can detect sub-proton masses[92], the flip of a single electronic spin[93], or the impact of a single photon[3]. Harnessing quantum mechanics in the operation of mechanical oscillators promises to enhance their performance, via *e.g.* quantum mechanical squeezing[94] or back action evasion[95], while also opening up new horizons in the fields of quantum computing and networking. Controlled single phonon excitations can form on-chip quantum transducers that mediate quantum information between remote or dissimilar elements, such as spins and photons or microwave and optical photons[96]. This provides several advantages over nano-photonic links such as tighter confinement, stronger couplings, slower speeds, and less crosstalk. Further, the long lifetimes of quantum mechanical excitations are promising for the storage of quantum information.

Mechanical resonators enter the quantum regime when the phonon occupation number, $n_m$ of a particular mode approaches 1, necessitating $k_B T/h f_m \to 1$ where $f_m$ is the mechanical oscillator's resonant frequency and $T$ is the oscillator mode's temperature. This requirement presents a significant challenge, as unlike optical photons, whose PHz-scale frequencies land them squarely in the quantum regime at room temperature, even the highest (~GHz-scale) frequency mechanical resonators see $n_m \gg 1$ above dilution refrigerator temperatures. Over the last decade, several different approaches have succeeded in overcoming this challenge, ranging from cooling in a cryogenic environment[91,97] to laser cooling[98,99]. However, it still imposes a significant technical hurdle to mechanical resonators' operation in the quantum regime. Also critical for observing quantum behaviour is a high mechanical quality factor $Q$, enabling a sufficiently low thermal decoherence rate $k_B T/\hbar Q$, which characterizes the inverse time for one quantum to enter from the environment when the oscillator is in the ground state. Another important figure of merit is the $Q f_m$ product, where $Q f_m > k_B T/\hbar$ is the requirement for neglecting thermal decoherence over one mechanical oscillation cycle. An intensive investigation into material properties, resonator geometries, and mechanical loss mechanisms has resulted in huge leaps forward in Q-factor engineering, with $Q$'s in excess of $5 \times 10^{10}$ and $Qf_m$ products in excess of $2 \times 10^{20}$ being realized[4].

Mechanical oscillators of many different forms, sizes, and materials have been explored to address the wide variety of end goals envisioned with operation in the quantum regime[3]. Several notable developments in the engineering of nanoresonators have significantly sped up progress towards harnessing quantum phenomena. As one example, the discovery that strained SiN membranes with sub micrometer-scale thicknesses can exhibit very large Q-factor [100] while being easily coupled to a macroscopic optical cavity has enabled laser cooling to the quantum mechanical ground state[99], observation of the quantum nature of the back action of light, see Fig. 5, and ponderomotive squeezing of light[101]. Optomechanical crystals[4] are another powerful nanoscale phonon-confining geometry that provides simultaneous strong confinement of high-Q, high frequency phononic modes and high-Q optical modes, as well as optomechanical coupling between the two DOF in a chip-scale integrated platform that can be realized in a variety of different materials. The high Qs arise from a separation in both real and frequency space of the resonator mode from other modes in the environment. The optomechanical coupling facilitates optical cooling and readout of the mechanics. Both SiN membranes and optomechanical resonators provide powerful platforms in which to realize quantum microwave-to-optical transduction, a key goal for quantum technologies, as well as long-lived mechanical quantum memories. Another exciting development in recent years is the idea of "soft-clamping", a technique that minimizes loss in stressed membrane-based resonators via dissipation dilution and strong mode confinement[102].

Mechanical oscillators make excellent sensors. Seminal experiments in 2004 used an ultrasensitive silicon nanocantilever to detect the magnetic field associated with a single electron spin via the dipolar magnetic force imparted to the cantilever[93], and experiments continue to march towards mechanical detection of single nuclear spins. In another remarkable feat, carbon nanotube mechanical resonators were used to detect a yoctogram change in mass, equivalent to that of a single proton, via changes in the resonator's frequency[92]. As mechanical oscillators are pushed to better sensitivities, the uncertainty principle of quantum mechanics looms large, imposing limits to their ultimate sensitivity. For instance, the so-called standard quantum limit (SQL) is an 'ordinary' limit that represents the optimal tradeoff between the precision of a measurement and the measurement's undesirable backaction. However, by clever quantum engineering, it is possible to break the SQL with approaches that include quantum nondemolition measurements[103] and utilization of quantum correlations in the noise. Using the latter technique, the SQL was recently experimentally broken in the measurement of a 20 nm thick SiN membrane resonator[104] by exploiting mechanically-induced quantum correlations in the spectrum of the light used to measure the oscillator. Another quantum-inspired sensitivity improvement invokes the concept of squeezing of the quantum mechanical noise. Squeezing of the mechanical motion of a micron-scale mechanical resonator coupled to a microwave circuit was demonstrated in 2015 [94].

Coupling mechanical oscillators to other quantum elements, such as spins[105] and atomic systems[106], is another exciting frontier that provides a means towards generating quantum nonlinearities and distinctly non-classical states in mechanical resonators[3]. Further mechanical coupling to spins could allow spin-driven cooling of the mechanics, mechanically-mediated squeezing of spins, and quantum transduction between localized spins and other quantum elements[107].

## Hybrid quantum systems

The previous sections gave an overview of how diverse degrees of freedom such as charge, spin, photons, and phonons are amenable to coherent quantum control and functionality at the nanoscale. Hybrid quantum systems[19, 108] constitute, in a sense, the culmination of the quantum-coherent nanoscience program – a platform where different DOF coherently interact with each other.

Transducing between DOF is often necessary to achieve basic functionalities such as measurement: for example, the quantum state of a single spin is extremely difficult to detect directly via its magnetic dipole[93]. Therefore, single-spin measurements require transduction to a charge for quantum dots[20], dopants in semiconductors[21], and surface atoms[6], or to a photon for optically-active defects[25,109], and molecules[34]. This transduction is accompanied by a large energy amplification, allowing detection with a classical apparatus, but destroys the coherence of the original quantum state.

Here we focus on the case where the coherence is preserved. This enables the fundamental functionalities of *transport* and *storage* of quantum states. Quantum state transport typically involves two or more DOF, one of which is tightly localised, while the other is more spread out or travels through a guiding structure. Quantum state storage serves the reverse purpose, either by locally storing the quantum state of a moving DOF or by swapping the state from an easily accessible DOF to a more isolated one, where a coherent superposition can be preserved for longer times.

The fundamental requirement for coherent hybrid systems is the existence of a coupling *g* between two DOF, with strength larger than the intrinsic dephasing rate *γ* of each DOF. Often, one of the DOF is of bosonic nature, such as a photon or a phonon, and is confined within a cavity. Here, the key parameter is the rate of loss or leakage $\kappa$ out of the cavity[110]. The regime of coherent hybridization is achieved when the cooperativity $C = g^2/\kappa\gamma$ exceeds unity.

The answer to `why nano?' is particularly subtle for hybrid quantum systems. The coupling $g$ is usually proportional to the relative volume shared by the different DOF. High $g$ can be achieved by confining to the nanoscale a bosonic mode to be coupled to an atomic-size object, or by increasing the footprint of the solid-state object, as in the case of superconducting qubits.

The development of circuit-quantum electrodynamics (cQED)[110, 111] gave a huge impetus to the field of hybrid quantum systems. The prototypical example is an engineered superconducting circuit whose charge state coherently couples to a photon in a microwave resonator. From the first Cooper-pair boxes coupled to planar resonators[112] to the millimetre-size transmons in bulk cavities[113], the coupling strengths and coherence times have steadily improved, often by increasing the physical size of the systems far beyond the nanoscale. Notably, recent work incorporating van der Waals heterostructures with temporally coherent cQED architectures seeks to reduce this footprint[114]. Nonetheless, the nonlinearity provided by a nanoscale Josephson junction remains the key ingredient for this technology.

Charge-photon hybridisation has also been achieved in semiconductor quantum dot systems. In the optical frequency range, strong coupling is routinely achieved between photons and atom-like levels in self-assembled quantum dots embedded in nanocavities[115] and, more recently, in SiV centres in diamond[79]. In the microwave regime, the size mismatch between the nanometre-scale of quantum dots and the millimetre-wavelength of the microwave photons requires exquisite material quality to minimize charge dephasing $\gamma$, and benefits from careful engineering of the microwave resonators[116], [117]. A high characteristic impedance $Z_r = \sqrt{\mathcal{L}/\mathcal{C}}$ of the resonator maximises the zero-point quantum fluctuations of the cavity voltage $V_{rms} \propto \omega_r/\sqrt{Z_r}$ and facilitates coupling to small electric dipoles. High-$Z_r$ is achieved by making very narrow and thin (down to $\sim 10$ nm) cavity wires to maximise the inductance[118], or by inserting SQUID loops[119].

Direct magnetic coupling of a single spin to a microwave photon is even more challenging and has not yet been achieved. This hurdle is circumvented by integrating nanoscale ferromagnets that produce intense magnetic field gradients, introducing a coherent spin-charge coupling for electrons in quantum dots[117,118], see Fig. 6. This results in a coherent spin-photon coupling, mediated by the charge, i.e. a three-DOF hybrid system. Even a four-DOF hybrid has been built[119], by connecting to the same microwave cavity a transmon and a spin qubit, the latter coupled to the cavity via its charge degree of freedom. A recent experiment demonstrated the detection of a single magnon in a ferromagnetic crystal by hybridising it with a transmon in a three-dimensional microwave cavity[120].

Building hybrid quantum systems requires balancing the contrasting requirements of high intrinsic coherence of the individual DOF with strong coupling between the DOF. Also, since the coupling is most often achieved when the DOF are in resonance, it is usually necessary to make at least one of the DOF individually tunable, which can open up channels for noise to disrupt the operation. The steady improvement in the cleanliness of materials and interfaces, and further developments in nanometre-scale lithography methods will bring an increasing number of physical systems within the realm of hybrid quantum systems.

## Conclusions and Outlook

In this Review, we have described the coherent quantum phenomena that arise in different DOF, such as charge, spin, electromagnetic radiation, and mechanical motion. We have emphasised the importance of controlling matter and waves at the nanometre scale in order to increase quantisation energies, interaction strengths and coherence times, and unlock novel functionalities explicitly enabled by quantum mechanics.

Paramount to any technological application of quantum-coherent nanoscience is the development of robust, reproducible, and scalable devices that can be built into larger-scale systems without compromising performance. Although we have organised our review along each type of DOF, the broader field of quantum coherent nanoscience can be further regrouped according to the manufacturing methods: "top-down" and "bottom-up".

In the top-down method, quantum coherent devices are produced using the tools of nanofabrication developed for the semiconductor and superconductor nanoelectronic industry, where feature sizes approach the 10 nm scale. This is the pathway used for gate-defined quantum dots, ion-implanted dopants, and superconducting circuits. Top-down fabrication allows natural integration with classical nanoelectronics for control and readout, and scalable manufacturing. Addressing an increasing number of qubits might require breaking with the two-dimensional, planar architecture and moving the interconnects into the third dimension[121]. The challenge for this approach will be to maintain uniformity and quantum coherence as the devices become more complex and the fabrication scales to ever smaller dimensions.

In the bottom-up method, quantum coherent devices are built upon the intrinsic quantum properties of atoms, molecules or point defects. At this atomic scale, the tools for top-down manufacturing can be ineffective because they are too coarse. Therefore, bottom-up manufacturing seeks to design components and interactions such that devices build themselves by directed self-assembly. Nanotechnologies building on the development of synthetic DNA[122] may offer the toolkit for these new construction paradigms. Integration of molecular qubits into such devices and electric field control[61, 123] of the qubits are key challenges. Atomic spin qubits on surfaces have the advantage that they can be arranged into arbitrary nanostructures with atomic-scale precision[124]. Most urgently, methods need to be developed to coherently control multiple coupled surface qubits. Finally, point defects in solids have excellent quantum coherence, extending even to room temperature. This enables great performance as quantum sensors. In such systems, one of the next challenges lies in creating controllable interactions between large arrays of such defects, for example through hybrid quantum devices linking these spin qubits to optical or mechanical DOF.

The coherent control of individual atoms and photons in vacuum[125] was the scientific breakthrough that led us into the second quantum revolution[126]. The scientific and technological program of quantum-coherent nanoscience consists of expanding this toolbox to harness quantum degrees of freedom within solid-state and molecular platforms built and controlled at the nanometre scale. In the coming years and decades, we expect quantum coherent nanoscience to make an impact in sensing and metrology applications, secure communications, quantum simulations, and universal quantum computing. Recent progress in this field has been tremendous and even cautious extrapolation into the near future promises exciting times with new discoveries and inventions.

**Fig.1 Ultrafast coherent control of two charge qubits in quantum dots**. **a.** False-colour scanning electron micrograph of the device. Plunger (P) and barrier (B) gates with P1 and P4 used for fast dc control of the quantum dots. $I_L$ ($I_R$) left (right) charge sensor current. **b.** Demonstration of the coherent evolution of the target qubit based on the state of the control qubit. Left: a one-dimensional slice through the data. Control and target qubits are controlled by pulses of varying length, $P_R^{Control}$ population of control (right) qubit. This work represents an important step towards electrical quantum control of two coupled qubits. Reprinted from ref.127.

**Fig 2. Quantum-coherent operation of two spin qubits in lithographically-defined nanostructures.** **a,** Schematic of a Si/SiGe double-quantum-dot device. Quantum dots D1 (purple circle) and D2 (orange circle) are used to confine two electron-spin qubits Q1 and Q2, respectively. P1 and P2 gate electrodes, MW1 and MW2 gate electrodes with microwave voltages, QW quantum well, 2DEG two-dimensional electron gas. **b,** Two-qubit density matrix (real part) demonstrating entanglement of two electron spins in neighbouring quantum dots. From the amplitude and sign of the non-zero entries, we see that the state is $(|00\rangle - |11\rangle)/\sqrt{2}$ with excellent fidelity. ρ density matrix, Re real part. **c,** Quantum control of individual P donors in silicon. False-coloured SEM image of a silicon nanoelectronic device with a single implanted phosphorous donor. The donor contains an electron (blue, with states $|\uparrow\rangle, |\downarrow\rangle$) and a nuclear (red, with states $|\Uparrow\rangle, |\Downarrow\rangle$) spin. **d,** After preparing the target state $(|\uparrow\Downarrow\rangle + |\downarrow\Uparrow\rangle)/\sqrt{2}$, measuring correlations between the electron and the nuclear state yields a Bell signal larger than 2, which implies that the spins must be quantum mechanically entangled. QND stands for quantum non-demolition measurement. **a** and **b** reprinted with permission from ref. 43, **c** and **d** from ref. 10.

**Fig. 3. Quantum-coherent control of the spin DOF of atomic and molecular spin bits. a,** The confined electric field in an STM tunnel junction can be used to coherently control the spin of a single atom on a surface. **b**, A spin-echo measurement reveals a $T_2$ time of 190ns measured on $S$ = ½ Ti spins on a thin MgO film grown on a metal. **c,** A single NV centre in a nanopillar of diamond is used as a quantum sensor. Optical excitation at 532nm, spin detection by monitoring the intensity of the lower-energy luminescence **d,** Dark contours in the NV magnetometry image of vortices in the superconductor $BaFe_2(As_{0.7}P_{0.3})_2$ correspond to locations where the stray field has a strength of 5.9 G. Scale bar, 400 nm. **e,** Tb phtalocyanine doppeldecker molecule trapped in a nanogap allows investigation of the electron-transport through the Pc ligand to control and monitor the electronic and nuclear magnetic

state of the Tb ion. VL (VR) left (right) electrode, VG gate electrode **f,** A quantum coherence time $T_2^*$ of 64 µs is obtained for the nuclear spin states of a single Tb ion in such a single-molecule device. **b** reprinted with permission from ref. 6, **c** and **d** from ref. 128, **f** from ref. 32.

**Fig. 4. Nanoscale Quantum Photonics**. **a**, Schematic of a Quantum dot (QD)-based single-photon experiment: The QD is inserted in a cryostat and cooled to 4 - 20 K. The QD emission is collected through a microscope objective that is also used for excitation, and then guided into a fibre. **b**, Schematic of a pillar microcavity used to improve the single-photon collection efficiency. A QD located at the centre of the pillar cavity experiences a Purcell-enhanced spontaneous emission rate. **c**, Schematic of a diamond photonic nanocavity containing two Si Vacancy (SiV) centres that are interacting through the cavity mode. **d**, Integrated nanophotonic device combining a single layer of InGaAs quantum dot (QD) single-photon emitters (shown in the AFM image on the top right), a 2 mm long GaAs ridge waveguide (WG), and an NbN superconducting single-photon detector (SSPD). **a** and **b** reproduced with permission from ref.76, **c** from ref.78, and **d** from ref.85.

**Fig. 5: Nanomechanical resonator at the quantum limit. a,** Schematic of an optomechanical cavity consisting of the optical mode of a Fabry-Perot cavity coupled to the motion of the $\omega_m = 2\pi \times 1.48$ MHz mode of a 40 nm thick $Si_3N_4$ drum resonator. Here, ppm refers to the transmission of the mirrors. **b,** Phonon occupation number of the membrane as a function of the cooling power of the laser. Dotted line is the limit imposed by the shotnoise of the laser. This work demonstrates sideband cooling of a micromechanical membrane resonator to the quantum backaction limit. Reprinted with permission from ref.99.

**Fig. 6: Photon – charge – spin hybrid quantum device.** A superconducting cavity, shown in panel **a**, confines a microwave photon such that the quantum fluctuations of the cavity voltage couple to the charge states of a semiconductor double quantum dot, shown in panel **b**. This results in the coherent hybridisation of the photon and charge states. Furthermore, a slanting magnetic field gradient introduces a strong coupling between the charge and spin degrees of freedom, resulting in the hybridisation of photon and spin states. Panel **c** illustrates the effect of charge-photon hybridisation in the frequency response of the cavity, which shows an anticrossing when the qubit and cavity states become resonant. This kind of hybrid quantum device relies crucially upon the nanoscale control of charge states in the quantum dot, and nanoscale magnetic field gradients. Reproduced with permission from ref.116.

**Box 1: Quantum coherence and the nanoscale**

Unlocking quantum coherent functionalities requires engineering quantum systems that possess discrete and sharp energy levels, while remaining accessible to the outside world for control and readout. Typical energy level spacings $hf$ are in the order of gigahertz (in frequency units). Since 1 GHz corresponds to a temperature of 50 mK, the quantum mechanical ground state can typically be achieved using dilution refrigerators. The ratio between $f$ (i.e. the resonance frequency) and dephasing rate $\gamma$ (i.e the resonance linewidth) is the coherence quality factor $Q$, which can exceed one million in the most performant systems. Charge and spin systems are often used as qubits for quantum information processing, where additional key parameters are the operation time $t_{op}$ required to coherently transition between quantum states, and the coherence time $T_2$ during which the phase of a quantum superposition state is preserved.

Almost all degrees of freedom in condensed matter systems require nanometre-scale dimensions in order to operate in the quantum coherent regime. Other dimensions of the system may be much larger, see e.g. superconducting devices which rely on a 1 nm tunnel barrier but have lateral dimensions of many micrometers[23].

The sketches represent **a**, scanning tunneling microscope performing spectroscopy of atoms on a surface. **b**, Josephson junction allowing coherent tunnelling of Cooper pairs. **c**, High-spin single-molecule magnet $Mn_{12}ac$. **d**, Lithographically-defined double quantum dot, confining electrons in a semiconductor. **e**, Self-assembled semiconductor quanutm dot, and its band diagram highlighting the energy of the emitted photons. **f**, Nanometre-thick mechanical resonator, allowing the control of quantized mechanical motion. Typical metrics for all systems are shown in the right column.

**Box 2: Methods to Coherently Drive Spins in Nanosystems**

Spins in nanosystems are well protected from the environment, which gives rise to relatively long quantum coherence times. The corollary is that it is challenging to coherently control and measure individual spins. The first coherent control methods borrowed the primary driving mechanism from ensemble spin resonance and utilized oscillating magnetic fields (upper row: S electron spin, $B_x$ magnetic field perpendicular to static field, $I_{RF}$ current at high angular frequency $\omega$). However, it is often desirable to confine the driving on the nanometer length scale, which requires electric fields.

Second row: converting an oscillating electric field into a magnetic field through the use of a permanent magnet, E electric field. Third row: Electric fields can also be used to modulate spin Hamiltonian parameters at high frequencies, HF hyperfine interaction, $V_g$ gate voltage, I nuclear spin. At much higher frequencies, optical transitions can be used for fast, coherent spin manipulation (lower row: $\Omega$ Rabi rates of two optical transitions, $\Delta$ detuning from third level, $\omega_1$ and $\omega_2$ energies of lower two levels). Diagram in third row reprinted with permission from ref. 32.

# References

1. Kastner, M. A. Artificial atoms. *Phys. Today* **46,** 24 (1993).
2. Joannopoulos, J. D., Villeneuve, P. R., & Fan, S. Photonic crystals. *Solid State Commun.* **102,** 165-173 (1997).
3. Aspelmeyer, M., Kippenberg, T. J. & Marquardt, F. Cavity optomechanics. *Rev. Mod. Phys.* **86,** 1391 (2014).
4. MacCabe, G. S. *et al.* Nano-acoustic resonator with ultralong phonon lifetime. *Science* **370,** 840 (2020).
5. Hayashi, T., Fujisawa, T., Cheong, H. D., Jeong, Y. H. & Hirayama, Y. Coherent manipulation of electronic states in a double quantum dot. *Phys. Rev. Lett.* **91,** 226804 (2003).
6. Yang, K. *et al.* Coherent spin manipulation of individual atoms on a surface. *Science* **366,** 509 (2019).
**First experimental work on the coherent manipulation of individual spins on a surface in scanning probe microscopy.**
7. He, Y. *et al.* A two-qubit gate between phosphorus donor electrons in silicon. *Nature* **571,** 371 (2019).
8. Ardavan, A. *et al.* Will spin-relaxation times in molecular magnets permit quantum information processing? *Phys. Rev. Lett.* **98,** 057201 (2007).
9. Dolde, F. *et al.* High-fidelity spin entanglement using optimal control. *Nat. Commun.* **5,** 3371 (2014).
10. Dehollain, J. P. *et al.* Bell's inequality violation with spins in silicon. *Nat. Nanotechnol.* **11,** 242 (2016).
11. Nichol, J. M. *et al.* High-fidelity entangling gate for double-quantum-dot spin qubits. *npj Quantum Inf.* **3,** 3 (2017).
12. Hanson, R., Kouwenhoven, L. P., Petta, J. R., Tarucha, S. & Vandersypen, L. M. K. Spins in few-electron quantum dots. *Rev. Mod. Phys.* **79,** 1217 (2007).
13. Chatterjee, A. *et al.* Semiconductor qubits in practice. *Nat. Rev. Phys.* **3,** 157 (2021).
14. Nakamura, Y., Chen C.D. & Tsai, J. S. Spectroscopy of energy-level splitting between two macroscopic quantum states of charge coherently superposed by Josephson coupling. *Phys. Rev. Lett.* **79,** 2328 (1997).
15. Bouchiat, V., Vion, D., Joyez, P., Esteve, D. & Devoret, M. H. Quantum coherence with a single Cooper pair. *Phys. Scr.* **76,** 165 (1998).
16. Nakamura, Y., Pashkin, Yu. A. & Tsai, J. S. Coherent control of macroscopic quantum states in a single-Cooper-pair box. *Nature* **398,** 786 (1999).
17. Zaretskey, F. V. *et al.* Decoherence in a pair of long-lived Cooper-pair boxes. *J. Appl. Phys.* **114,** 094305 (2013).
18. Rabl, P. *et al.* A quantum spin transducer based on nanoelectromechanical resonator arrays. *Nat. Phys.* **6,** 602 (2010).
19. Kurizki, G. *et al.* Quantum technologies with hybrid systems. *Proc. Nat. Acad. Sci*. **112,** 3866 (2015).
20. Elzerman, J. M. *et al.* Single-shot read-out of an individual electron spin in a quantum dot. *Nature* **430,** 431-435 (2004).
**The ability to perform projective quantum measurement of a single electron spin by electrical means opened the door to the practical use of spins in semiconductor quantum devices**
21. Morello, A. *et al.* Single-shot readout of an electron spin in silicon. *Nature* **467,** 687 (2010).
22. Koch, J. *et al.* Charge-insensitive qubit design derived from the Cooper pair box*. Phys. Rev. A* **76,** 042319 (2007).
23. Krantz, P. *et al.* A Quantum Engineer's Guide to Superconducting Qubits*. Appl. Phys. Rev.* **6,** 021318 (2019).

24. Arute, F. *et al.* Quantum supremacy using a programmable superconducting processor. *Nature* **574,** 505 (2019).
25. Jelezko, F., Gaebel, T., Popa, I., Gruber, A. & Wrachtrup, J. Observation of coherent oscillations in a single electron spin. *Phys. Rev. Lett*. **92,** 076401 (2004).
26. Wu, Y., Wang, Y., Qin, X., Rong, X. & Du, J. A programmable two-qubit solid-state quantum processor under ambient conditions. *npj Quantum Inf.* **5,** 9 (2019).
27. Watson, T. F. *et al.* Atomically engineered electron spin lifetimes of 30 s in silicon. *Sci. Adv.* **3,** e1602811 (2017).
28. Anderson, C. P. *et al.* Electrical and optical control of single spins integrated in scalable semiconductor devices. *Science* **366,** 1225 (2020).
29. Loss, D. & DiVincenzo, D. P. Quantum computation with quantum dots. *Phys. Rev. A* **57,** 120 (1998).
30. Zwanenburg, F. A. *et al.* Silicon quantum electronics. *Rev. Mod. Phys.* **85,** 961 (2013).
31. Vandersypen, L. M. K. & Eriksson, M. A. Quantum computing with semiconductor spins. *Phys. Today* **72 (8),** 38 (2019).
32. Thiele, S. *et al.* Electrically driven nuclear spin resonance in single-molecule magnets. *Science,* **344,** 1135 (2014).
**Quantum-coherent control of an individual molecular spin in an electronic device.**
33. Malavolti, L. *et al.* Tunable spin–superconductor coupling of spin 1/2 vanadyl phthalocyanine molecules. *Nano Lett.* **18,** 7955-7961 (2018).
34. Bayliss S. L. *et al.* Optically addressable molecular spins for quantum information processing. *Science* **370,** 1309 (2020).
35. Baumann, S. *et al.* Electron paramagnetic resonance of individual atoms on a surface. *Science* **350,** 417 (2015).
36. Seifert, T. S. *et al.* Single-atom electron paramagnetic resonance in a scanning tunneling microscope driven by a radio-frequency antenna at 4 K. *Phys. Rev. Res.* **2,** 013032 (2020).
37. Yale, C.G. *et al.* All-optical control of a solid-state spin using coherent dark states. *PNAS* **110**, 7595 (2013).
38. Awschalom, D. D., Hanson, R., Wrachtrup, J. & Zhou, B. B. Quantum technologies with optically interfaced solid-state spins. *Nat. Photonics* **12,** 516 (2018).
39. Petta, J. R. *et al.* Coherent Manipulation of Coupled Electron Spins in Semiconductor Quantum Dots. *Science* **309**, 2180 (2005).
40. K. Eng *et al.* Isotopically enhanced triple-quantum-dot qubit. *Science Advances* **1**:e1500214 (2015)
41. Veldhorst, M. *et al.* An addressable quantum dot qubit with fault-tolerant control-fidelity. *Nat. Nanotechnol.* **9,** 981 (2014).
42. Veldhorst, M. *et al.* A two-qubit logic gate in silicon. *Nature* **526**, 410 (2015).
**First experimental demonstration of two-qubit logic operations in silicon, the same platform used for classical nanoelectronics**
43. Watson, T. F. *et al.* A programmable two-qubit quantum processor in silicon. *Nature* **555,** 633-637 (2018).
44. Mills, A. R. *et al.* Shuttling a single charge across a one-dimensional array of silicon quantum dots. *Nat. Commun.* **10,** 1063 (2019).
45. Xue, X. *et al.* Computing with spin qubits at the surface code error threshold. https://arxiv.org/abs/2107.00628 (2021)
46. K. Takeda *et al*. Quantum tomography of an entangled three-qubit state in silicon. *Nature Nanotech.* (2021), https://doi.org/10.1038/s41565-021-00925-0
47. Hendrickx, N. W. *et al.* A four-qubit germanium quantum processor. *Nature* **591,** 580–585 (2021).
48. Saeedi, K. *et al.* Room-temperature quantum bit storage exceeding 39 minutes using ionized donors in silicon-28. *Science* **342,** 830 (2013).

49. Muhonen, J. T. *et al.* Storing quantum information for 30 seconds in a nanoelectronic device. *Nat. Nanotechnol.* **9**, 986 (2014).
50. Mądzik, M. T. *et al.* Precision tomography of a three-qubit electron-nuclear quantum processor in silicon. https://arxiv.org/abs/2106.03082 (2021)
51. Mądzik, M. T. *et al.* Conditional quantum operation of two exchange-coupled single-donor spin qubits in a MOS-compatible silicon device. *Nat. Commun.* **12,** (2020).
52. Gruber, A. *et al.* Scanning confocal optical microscopy and magnetic resonance on single defect centers. *Science* **276,** 2012 (1997).
53. Bradley, C. E. *et al.* A Ten-qubit solid-state spin register with quantum memory up to one minute. *Phys. Rev. X* **9,** 031045 (2019).
54. Myers, B. A. *et al.* Probing Surface Noise with Depth-Calibrated Spins in Diamond. *Phys. Rev. Lett.* **113,** 027602 (2014).
55. Smith, J. M., Meynell, S. A., Bleszynski Jayich, A. C. and Meijer, J. Colour centre generation in diamond for quantum technologies. *Nanophotonics* **8**, 1889 (2019).
56. Lado, J. L., Ferrón, A. & Fernández-Rossier, J. Exchange mechanism for electron paramagnetic resonance of individual adatoms. *Phys. Rev. B* **96,** 205420 (2017).
57. Willke, P. *et al.* Probing Quantum Coherence in Single Atom Electron Spin Resonance", *Science Advances* **4**, eaaq1543 (2018).
58. Leuenberger, M. N. & Loss, D. Quantum computing in molecular magnets. *Nature,* **410,** 789-793 (2001).
59. Zadrozny, J. M., Niklas, J., Poluektov, O. G.& Freedman, D. E. Millisecond coherence time in a tunable molecular electronic spin qubit. *ACS Cent. Sci.* **1,** 488 (2015).
60. Atzori, M. *et al.* Room-temperature quantum coherence and Rabi oscillations in vanadyl phthalocyanine: Toward nultifunctional molecular spin qubits. *J. Am. Chem. Soc.* **138,** 2154 (2016).
61. Liu, J, *et al*. Quantum coherent spin-electric control in a molecular nanomagnet at clock transitions. accepted at *Nat. Phys.* (2021), https://arxiv.org/abs/2005.01029
62. Moreno-Pineda, E. & Wernsdorfer, W. Measuring molecular magnets for quantum technologies, *Nat. Rev. Phys.* (2021). https://doi.org/10.1038/s42254-021-00340-3
63. Zhou, Y., Kanoda, K. & Ng, T.-K. Quantum spin liquid states. *Rev. Mod. Phys.* **89,** 025003 (2017).
64. Gross, C. & Bloch, I. Quantum simulations with ultracold atoms in optical lattices. *Science* **357,** 995 (2017).
65. Choi, D. *et al.* Colloquium: Atomic spin chains on surfaces. *Rev. Mod. Phys.* **91,** 041001 (2019).
66. Tacchino, F., Chiesa, A., Carretta, S. & Gerace, D. Quantum computers as universal quantum simulators: State-of-the-art and perspectives. *Adv. Quantum Technol.* **3,** 1900052 (2020).
67. Salfi, J. *et al.* Quantum simulation of the Hubbard model with dopant atoms in silicon. *Nat. Commun.* **7,** 1 (2016).
68. Yang, K. *et al.* Probing resonating valence bond states in artificial quantum magnets. *Nat. Commun.* **12,** 993 (2021).
69. Dehollain, J. P. *et al.* Nagaoka ferromagnetism observed in a quantum dot plaquette. *Nature* **579,** 528 (2020).
70. Flamini, F., Spagnolo, N. & Sciarrino, F. Photonic quantum information processing: a review. *Rep. Prog. Phys.* **82,** 016001 (2018).
71. Wehner, S., Elkouss, D. & Hanson, R. Quantum internet: A vision for the road ahead. *Science* **362,** eaam9288 (2018).
72. Wan, N. H. *et al.* Large-scale integration of artificial atoms in hybrid photonic circuits. *Nature* **583,** 226 (2020).
**Demonstration of state-of-the-art photonic circuits constructed by placing quantum microchips with diamond colour centres on top of aluminium nitride photonic waveguides.**

73. Reiserer, A. & Gerhard Rempe, G. Cavity-based quantum networks with single atoms and optical photons. *Rev. Mod. Phys.* **87**, 1379 (2015).
74. Wada, O. Femtosecond all-optical devices for ultrafast communication and signal processing. *New J. of Phys.* **6,** 183 (2004).
75. Volz, T. *et al.* Ultrafast all-optical switching by single photons. *Nat. Photonics* **6,** 605 (2012).
76. Senellart, P., Solomon, G. & White, A. High-performance semiconductor quantum-dot single-photon sources. *Nat. Nanotechnol.* **12,** 1026 (2017).
77. You, L. Superconducting Nanowire Single-Photon Detectors for Quantum Information. *Nanophotonics* **9**, 2673 (2020).
78. Evans, R. E. *et al.* Photon-mediated interactions between quantum emitters in a diamond nanocavity. *Science* **362,** 662 (2018).
79. Bhaskar, M. K. *et al.* Experimental demonstration of memory-enhanced quantum communication. *Nature* **580,** 60 (2020).
80. D'Amico, I. *et al.* Nanoscale quantum optics. *Riv. Nuovo Cimento* **4,** 153 (2019).
81. Aharonovich, I., Englund, D. & Toth, M. Solid-state single-photon emitters. *Nat. Photonics* **10,** 631 (2016).
82. Grosso, G. *et al.* Tunable and high-purity room temperature single-photon emission from atomic defects in hexagonal boron nitride. *Nat. Commun.* **8,** 705 (2017).
83. Etzelmüller Bathen, M. & Vines, L. Manipulating Single-Photon Emission from Point Defects in Diamond and Silicon Carbide", *Adv. Quantum Technol.*, 2100003 (2021).
84. Badolato, A. *et al.* Deterministic coupling of single quantum dots to single nanocavity modes. *Science* **308,** 1158 (2005).
85. Reithmaier, G. *et al.* On-chip generation, routing, and detection of resonance fluorescence. *Nano Lett.* **15,** 5208 (2015).
86. Kavokin, A., Baumberg, J. J., Malpuech, G. & Laussy, F. P. Microcavities. *Oxford University Press* (2017).
87. Krauss, T. F. Why do we need slow light? *Nat. Photonics* **2,** 448 (2008).
88. Burkard, G., Gullans, M. J., Mi, X. & Petta. J. R. Superconductor–semiconductor hybrid-circuit quantum electrodynamics. *Nat. Rev. Phys.* **2,** 129 (2020).
89. Mamin, H. J. & Rugar, D. Sub-attonewton force detection at millikelvin temperatures. *Appl. Phys. Lett.* **79**, 3358 (2001).
90. Weber, P. *et al.* Force sensitivity of multilayer graphene optomechanical devices. *Nat. Commun.* **7**, 12496 (2016).
91. Fogliano F. *et al.* Ultrasensitive nano-optomechanical force sensor operated at dilution temperatures, *Nat. Commun.* **12**, 4124 (2021).
92. Chaste, J. *et al.* A nanomechanical mass sensor with yoctogram resolution. *Nat. Nanotechnol.* **7,** 301–304 (2012).
93. Rugar, D., Budakian, R., Mamin, H. J. & Chui B. W. Single spin detection by magnetic resonance force microscopy. *Nature* **430,** 329 (2004).
**Breakthrough experimental results on measuring the dipolar magnetic force from a single electron spin.**
94. Wollman, E. E., Lei, C. U., Weinstein, A. J. & Suh, J. Quantum squeezing of motion in a mechanical resonator. *Science* **349,** 952 (2015).
95. Shomroni, I., Qiu, L., Malz, D., Nunnenkamp, A. & Kippenberg, T. J. Optical backaction-evading measurement of a mechanical oscillator. *Nat. Commun.* **10,** 2086 (2019).
96. Wu M., Zeuthen E., Balram K.C. , Srinivasan K. Microwave-to-Optical Transduction Using a Mechanical Supermode for Coupling Piezoelectric and Optomechanical Resonators. *Phys. Rev. Applied* **13**, 014027 (2020).

97 O'Connell, A. D. *et al.* Quantum ground state and single-phonon control of a mechanical resonator. *Nature* **464,** 697 (*2010*).
**Control of mechanical motion down to the last quantum of excitation, in a nanostructured mechanical oscillator**

98 Chan, J. *et al*. Laser cooling of a nanomechanical oscillator into its quantum ground state. *Nature* **478,** 89 *(*2011*)*.

99 R. W. Peterson, T. P. Purdy, N. S. Kampel, R. W. Andrews, P.-L. Yu, K. W. Lehnert, and C. A. Regal, "Laser Cooling of a Micromechanical Membrane to the Quantum Backaction Limit", *Phys. Rev. Lett.* **116**, 063601 (2016).

100 Zwickl, B. M. *et al.* High quality mechanical and optical properties of commercial silicon nitride membranes. *Appl. Phys. Lett.* **92,** 103125 (2008).

101 Purdy, T. P., Yu, P.-L., Peterson, R. W., Kampel, N. S. & Rega, C. A. Strong optomechanical squeezing of light. *Phys. Rev. X* **3,** 031012 (2013).

102 Tsaturyan, Y., Barg, A., Polzik, E. S. & Schliesser, A. Ultracoherent nanomechanical resonators via soft clamping and dissipation dilution. *Nat. Nanotechnol.* **12,** 776 (2017).

103 Braginsky, V.B., Khalili, F.Y. Quantum Measurement. Cambridge Univ. Press, Cambridge, (1992).

104 Mason, D., Chen, J., Rossi, M., Tsaturyan, Y. & Schliesser, A. Continuous force and displacement measurement below the standard quantum limit. *Nat. Phys.* **15,** 745 (2019).

105 Ganzhorn, M., Klyatskaya, S., Ruben, M. & Wernsdorfer, W. Strong spin–phonon coupling between a single-molecule magnet and a carbon nanotube nanoelectromechanical system. *Nat. Nanotechnol.* **8,** 165 (2013).

106 Karg, T. M. *et al.* Light-mediated strong coupling between a mechanical oscillator and atomic spins 1 meter apart. *Science* **369,** 174 (2020).

107 Lee, D., Lee, K. W., Cady, J. V., Ovartchaiyapong, P. & Jayich, A. C. B. Topical Review: Spins and mechanics in diamond. *J. Opt.* **19,** 033001 (2017).

108 Xiang, Z., Ashhab, S., You, J. Q. & Nori, F. Hybrid quantum circuits: Superconducting circuits interacting with other quantum systems. *Rev. Mod. Phys*. **85,** 623 (2013).

109 Robledo, L. *et al.* High-fidelity projective read-out of a solid-state spin quantum register. *Nature* **477,** 574 (2011).

110 Blais, A., Huang, R. S., Wallraff, A., Girvin, S. M. & Schoelkopf, R. J. Cavity quantum electrodynamics for superconducting electrical circuits: An architecture for quantum computation. *Phys. Rev. A* **69,** 062320 (2004).

111 Blais, A., Girvin, S. M. & Oliver, W.D. Quantum information processing and quantum optics with circuit quantum electrodynamics. *Nat. Phys.* **16,** 247-256 (2020).

112 Wallraff, A. *et al.* Strong coupling of a single photon to a superconducting qubit using circuit quantum electrodynamics. *Nature* **431,** 162 (2004).
**Demonstration of strong coupling between a microwave photon and a superconducting circuit, enabling the hybridisation of two disparate quantum systems**

113 Paik, H. *et al.* Observation of high coherence in Josephson junction qubits measured in a three-dimensional circuit QED architecture. *Phys. Rev. Lett*. **107,** 240501 (2011).

114 Wang, J. I-J. *et al.* Coherent control of a hybrid superconducting circuit made with graphene-based van der Waals heterostructures. *Nat. Nanotechnol.* **14,** 120-125 (2019).

115 Yoshie, T. *et al.* Vacuum Rabi splitting with a single quantum dot in a photonic crystal nanocavity. *Nature* **432,** 200 (2004).

116 Mi, X., Cady, J. V., Zajac, D. M., Deelman, P. W. & Petta, J. R. Strong coupling of a single electron in silicon to a microwave photon. *Science* **355,** 156 (2017).

117 Mi, X. *et al.* A Coherent spin-photon interface in silicon. *Nature* **555**, 559 (2018).

118 Samkharadze, N. *et al.* Strong spin-photon coupling in silicon. *Science* **359,** 1123 (2018).

**Refs. 117 and 118 demonstrate hybrid quantum nanoelectronic devices, where an electron spin coherently couples to a microwave photon via the electron's charge**

[119] Landig, A. J. *et al.* Virtual-photon-mediated spin-qubit–transmon coupling. *Nat. Commun.* **10,** 5037 (2019).

[120] Lachance-Quirion. D. *et al.* Entanglement-based single-shot detection of a single magnon with a superconducting qubit. *Science* **367,** 425 (2020).

[121] Rosenberg, D. *et al.* 3D integration and packaging for solid-state qubits. *IEEE Microw. Mag.* **21 (8),** 72-86 (2020).

[122] Rothemund, P. W. K. Folding DNA to create nanoscale shapes and patterns. *Nature* **440,** 297 (2006).

[123] Fittipaldi, M. *et al.* Electric field modulation of magnetic exchange in molecular helices. *Nat. Mater.* **18**, 329 (2019).

[124] Eigler, D. M. & Schweizer, E. K. Positioning single atoms with a scanning tunnelling microscope. *Nature* **344,** 524 (1990).

[125] Wineland, D. J. Nobel Lecture: Superposition, entanglement, and raising Schrödinger's cat. *Rev. Mod. Phys.* **85**, 1103 (2013).

[126] Dowling, J. P. & Gerard J. Milburn, G. J. Quantum technology: the second quantum revolution", *Philosoph. Trans. of the Royal Society* **A 361**, 1227 (2003).

[127] MacQuarrie, E. R. *et al.* Progress toward a capacitively mediated CNOT between two charge qubits in Si/SiGe. *npj Quantum Inf.* **6,** 81 (2020).

[128] Pelliccione, M. *et al.* Scanned probe imaging of nanoscale magnetism at cryogenic temperatures with a single-spin quantum sensor. *Nat. Nanotechnol.* **11,** 700 (2016).

[129] Wilkinson, T. A. *et al.* Spin-selective AC Stark shifts in a charged quantum dot. *Appl. Phys. Lett.* **114**, 133104 (2019).

[130] Press, D. *et al.* Complete quantum control of a single quantum dot spin using ultrafast optical pulses. *Nature* **456**, 218 (2008).

[131] Buckley, B.B, Fuchs, G.D., Bassett, L. C. & Awschalom, D. D. Spin-Light Coherence for Single-Spin Measurement and Control in Diamond. *Science* **330**, 1212 (2010).

[132] Tamarat, Ph. Spin-flip and spin-conserving optical transitions of the nitrogen-vacancy centre in diamond. *New J. Phys.* **10**, 045004 (2008).

[133] Zhong, M. *et al.* Optically addressable nuclear spins in a solid with a six-hour coherence time. *Nature* **517**, 177 (2015).

## Acknowledgements

A.J.H. acknowledges financial support from the Institute for Basic Science under IBS-R027-D1. W.D.O. from the U.S. Army Research Office grant number W911WF-18-1-0116 and the National Science Foundation grant number PHY-1720311. L.M.K.V. from the European Research Council (grant no. 882848), A.A. from the UK Engineering and Physical Sciences Research Council (EP/P000479/1), and the European Union's Horizon 2020 research and innovation programme under grant agreement numbers 863098 and 862893. R.S. from EU-H2020 research project 862893. A. B. J. from the NSF award QIS-1820938, and the NSF QLCI through grant number OMA-2016245. J.F.R. from Generalitat Valenciana funding Prometeo 2017/139 and MINECO-Spain (Grant No. PID2019-109539GB), A.L. from the UNSW Scientia Program, and A.M. from the Australian Research Council (CE170100012 and DP180100969), the US Army Research Office (grant number W911NF-17-1-0200), and the Australian Department of Industry, Innovation and Science (AUSMURI00002). The views and conclusions contained in this document are those of the authors and should not be interpreted as representing the official policies, either expressed or implied, of the ARO or the US Government. The US Government is authorized to reproduce and distribute reprints for government purposes notwithstanding any copyright notation herein.

## Ethics Declarations

The authors declare no competing interests.

## Additional Information

Reprints and permission information is available online at www.nature.com/reprints. Correspondence and requests for materials should be addressed to AJH or AM.

**Box 1**

| | | | |
|---|---|---|---|
| charge | 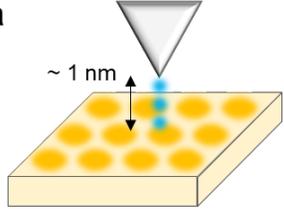 **a** ~ 1 nm | **b** ~ 1 nm | $f \sim 1 - 10$ GHz<br>$Q \sim 10^4 - 10^7$<br>$t_{op} \sim 0.1 - 100$ ns<br>$T_2 \sim 1 - 100$ μs (superconductors)<br>$\phantom{T_2 \sim}\ 1 - 100$ ns (semiconductors) |
| spin | **c** ~ 2 nm | **d** ~ 50 nm | $f \sim 1 - 50$ GHz (electrons)<br>$\phantom{f \sim}\ 1 - 100$ MHz (nuclei)<br>$Q \sim 10^4 - 10^7$<br>$t_{op} \sim 10 - 1000$ ns (electrons)<br>$\phantom{t_{op} \sim}\ 1 - 100$ μs (nuclei)<br>$T_2 \sim 100$ ns $- 1$ s (electrons)<br>$\phantom{T_2 \sim}\ 0.1$ ms $- 100$ s (nuclei) |
| photons | **e** ~ 20 nm | $E$, $E_C$, $E_V$ | $f \sim 10^5 - 10^6$ GHz<br>$Q \sim 10^5 - 10^6$ |
| mechanics | **f** ~ 100 nm | | $f \sim 0.01 - 5$ GHz<br>$Q \sim 10^6 - 10^{10}$ |

**Box 2**

| Time-dependent drive | Examples of corresponding time-dependent Hamiltonian | Schematic of an example | Some experimental systems and reference papers exploiting this mechanism |
|---|---|---|---|
| Directly applied magnetic field, $B_x$ along $x$ | $B_x \cos(\omega t)\, \boldsymbol{S}_x$ | 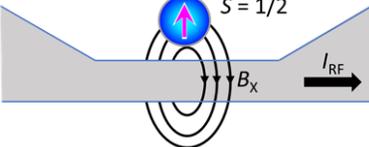 | - Magnetic defects in insulators and semiconductors [21, 25]<br>- Ensemble ESR and NMR [8]<br>- Lithographic quantum dots [12, 31] |
| Electric-field-induced displacement in an inhomogeneous magnetic field, $B_x(x)$ | $\dfrac{\partial B}{\partial x}\dfrac{\partial x}{\partial E} E \cos(\omega t)\, \boldsymbol{S}_x$ | 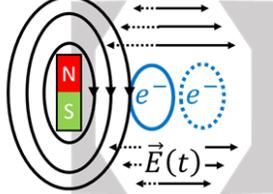 | - Lithographic quantum dots [31]<br>- Magnetic atoms on surfaces in STM [6, 36, 56] |
| Electric-field-modulated spin-Hamiltonian Parameter | Hyperfine interaction:<br>$A(E)\, \vec{\boldsymbol{S}} \cdot \vec{\boldsymbol{I}}$<br>Exchange interaction:<br>$J(E)\, \vec{\boldsymbol{S_1}} \cdot \vec{\boldsymbol{S_2}}$<br>Anisotropy energy:<br>$D(E)\, \boldsymbol{S_z}^2$ | 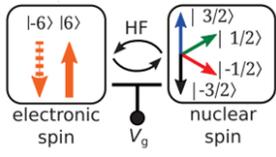 | - Nuclear spin control in individual magnetic molecules in junction [32]<br>- Lithographic quantum dots [12, 31, 39]<br>- High-spin magnetic atoms on surfaces in STM [35, 36] |
| Optical spin control | $\hbar \begin{pmatrix} 0 & 0 & \dfrac{-\Omega_{1,3}}{2} \\ 0 & \omega_2 - \omega_1 & \dfrac{-\Omega_{2,3}}{2} \\ \dfrac{-\Omega^*_{1,3}}{2} & \dfrac{-\Omega^*_{2,3}}{2} & \Delta \end{pmatrix}$ | 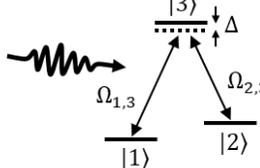 | - Self-assembled quantum dots [129, 130]<br>- Colour centres [131, 132]<br>- Lanthanide spins in insulators [133] |

**Figure 1**

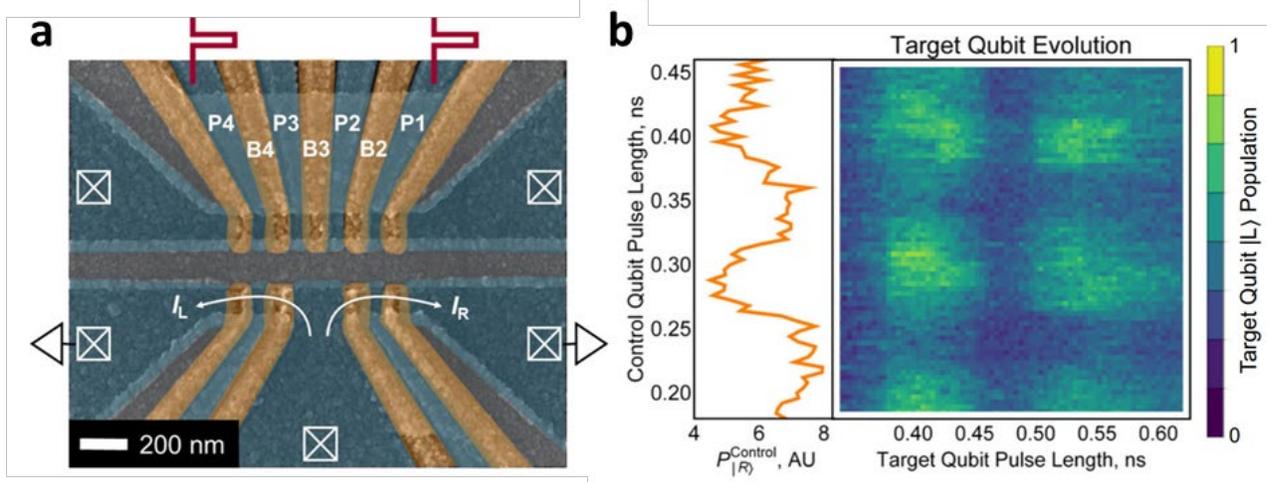

**Figure 2**

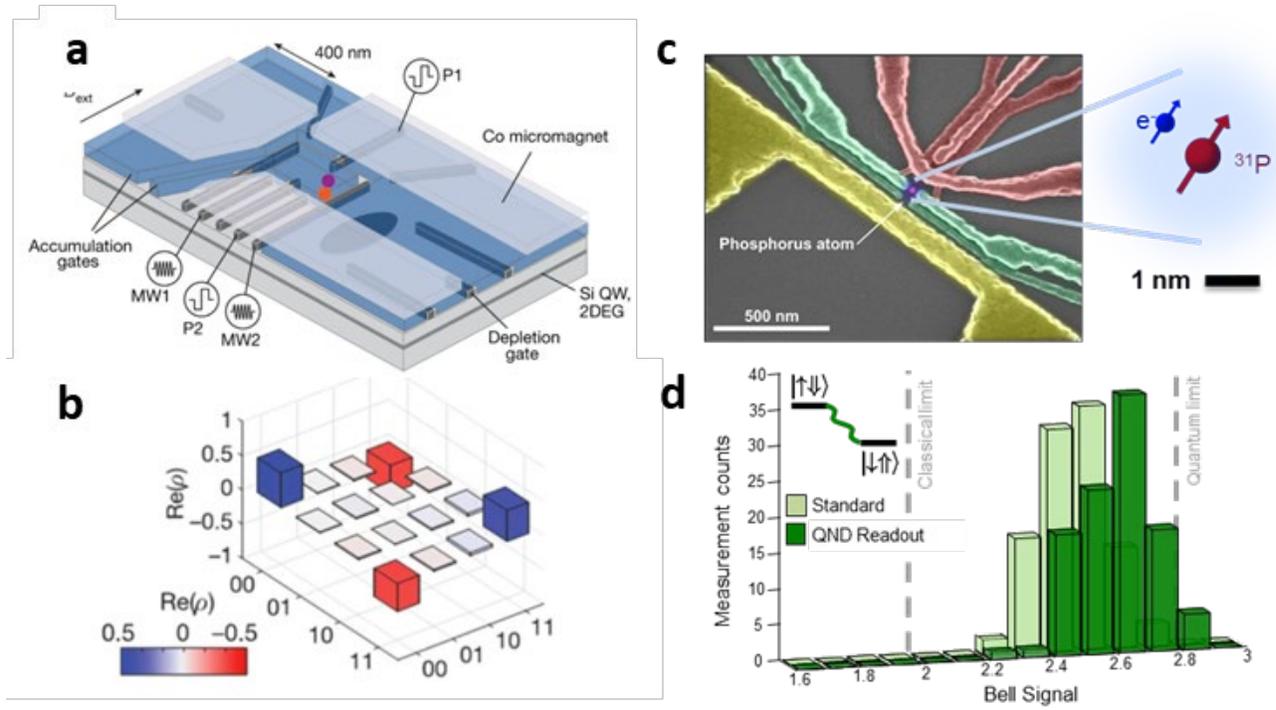

**Figure 3**

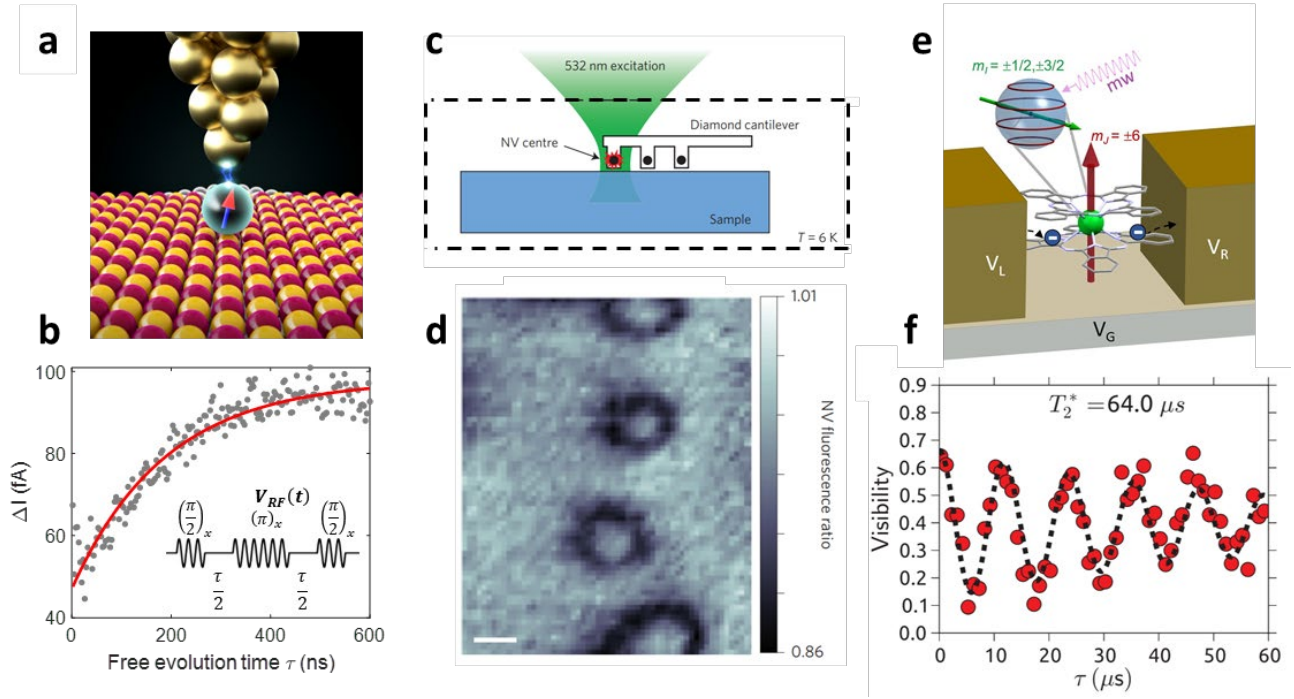

**Figure 4**

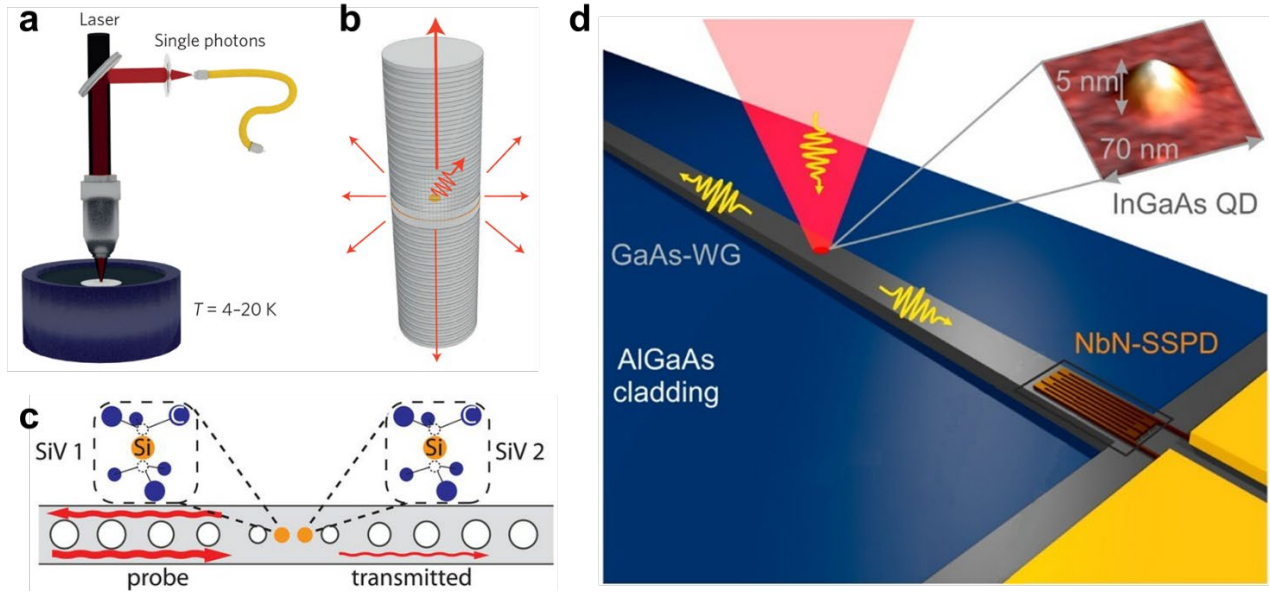

**Figure 5**

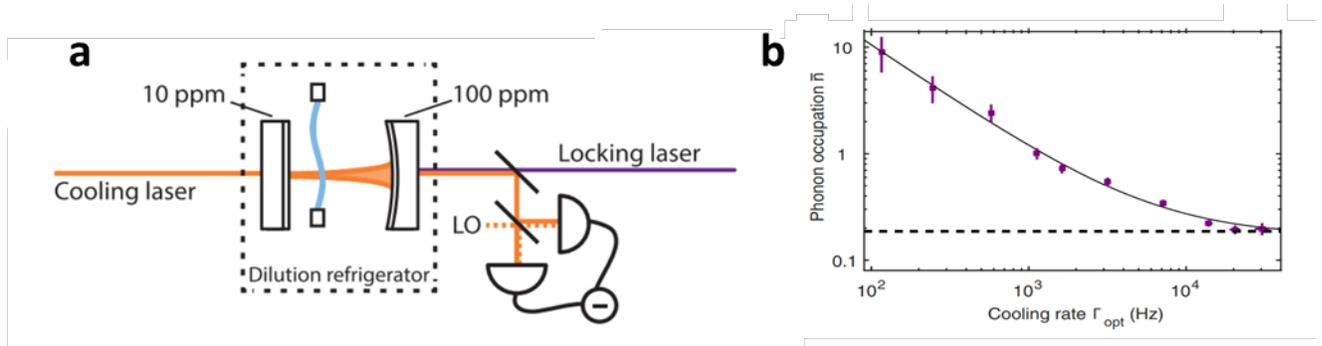

**Figure 6**

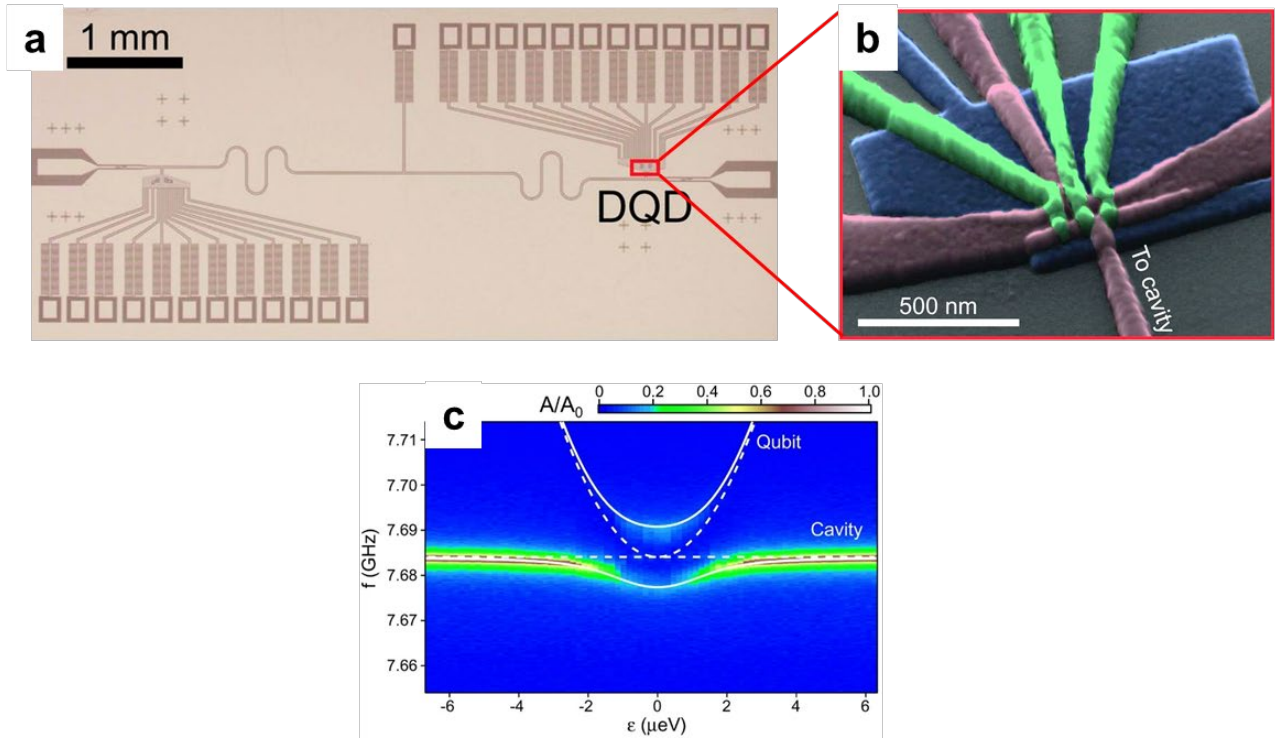